\documentclass[acmtog,nonacm]{acmart}

\usepackage{booktabs} 
\usepackage{rotating}

\usepackage{multirow}
\usepackage[ruled]{algorithm2e} 

\usepackage[normalem]{ulem} 
\usepackage[export]{adjustbox}
\SetAlFnt{\small}
\SetAlCapFnt{\small}
\SetAlCapNameFnt{\small}
\SetAlCapHSkip{0pt}
\newcommand{\vcentered}[1]{\raisebox{1.75\height}{#1}} 

\copyrightyear{2024}
\acmYear{2024}
\setcopyright{acmlicensed}\acmConference[SIGGRAPH Conference Papers '24]{Special Interest Group on Computer Graphics and Interactive Techniques Conference Conference Papers '24}{July 27-August 1, 2024}{Denver, CO, USA}
\acmBooktitle{Special Interest Group on Computer Graphics and Interactive Techniques Conference Conference Papers '24 (SIGGRAPH Conference Papers '24), July 27-August 1, 2024, Denver, CO, USA}
\acmDOI{10.1145/3641519.3657444}
\acmISBN{979-8-4007-0525-0/24/07}

\begin{document}
\title{TexSliders: Diffusion-Based Texture Editing in CLIP Space}

\author{Julia Guerrero-Viu}
\affiliation{%
	\institution{Universidad de Zaragoza - I3A}
	\country{Spain}
}
\email{juliagviu@unizar.es}
\orcid{0000-0002-2077-683X}

\author{Milos Hasan}
\affiliation{%
	\institution{Adobe Research}
	\country{USA}
}
\email{mihasan@adobe.com}
\orcid{0000-0003-3808-6092}

\author{Arthur Roullier}
\affiliation{%
	\institution{Adobe Research}
	\country{France}
}
\email{roullier@adobe.com}
\orcid{0000-0002-6483-240X}

\author{Midhun Harikumar}
\affiliation{%
	\institution{Adobe}
	\country{USA}
}
\email{mharikum@adobe.com}
\orcid{0009-0009-2199-5063}

\author{Yiwei Hu}
\affiliation{%
	\institution{Adobe Research}
	\country{USA}
}
\email{yiwhu@adobe.com}
\orcid{0000-0002-3674-295X}

\author{Paul Guerrero}
\affiliation{%
	\institution{Adobe Research}
	\country{United Kingdom}
}
\email{guerrero@adobe.com}
\orcid{0000-0002-7568-2849}

\author{Diego Gutierrez}
\affiliation{%
	\institution{Universidad de Zaragoza - I3A}
	\country{Spain}
}
\email{diegog@unizar.es}
\orcid{0000-0002-7503-7022}

\author{Belen Masia}
\affiliation{%
	\institution{Universidad de Zaragoza - I3A}
	\country{Spain}
}
\email{bmasia@unizar.es}
\orcid{0000-0003-0060-7278}

\author{Valentin Deschaintre}
\affiliation{%
	\institution{Adobe Research}
	\country{United Kingdom}
}
 \email{deschain@adobe.com}
 \orcid{0000-0002-6219-3747}

\begin{abstract}
Generative models have enabled intuitive image creation and manipulation using natural language. In particular, diffusion models have recently shown remarkable results for natural image editing. In this work, we propose to apply diffusion techniques to edit \emph{textures}, a specific class of images that are an essential part of 3D content creation pipelines. We analyze existing editing methods and show that they are not directly applicable to textures, since their common underlying approach, manipulating attention maps, is unsuitable for the texture domain. To address this, we propose a novel approach that instead manipulates CLIP image embeddings to condition the diffusion generation.
We define editing directions using simple text prompts (e.g., ``aged wood'' to ``new wood'') and map these to CLIP image embedding space using a texture prior, with a sampling-based approach that gives us identity-preserving directions in CLIP space. To further improve identity preservation, we project these directions to a CLIP subspace that minimizes identity variations resulting from entangled texture attributes. Our editing pipeline facilitates the creation of arbitrary sliders using natural language prompts only, with no ground-truth annotated data necessary.
\end{abstract}

%
%



\begin{teaserfigure}
\centering
 \includegraphics[width=\textwidth]{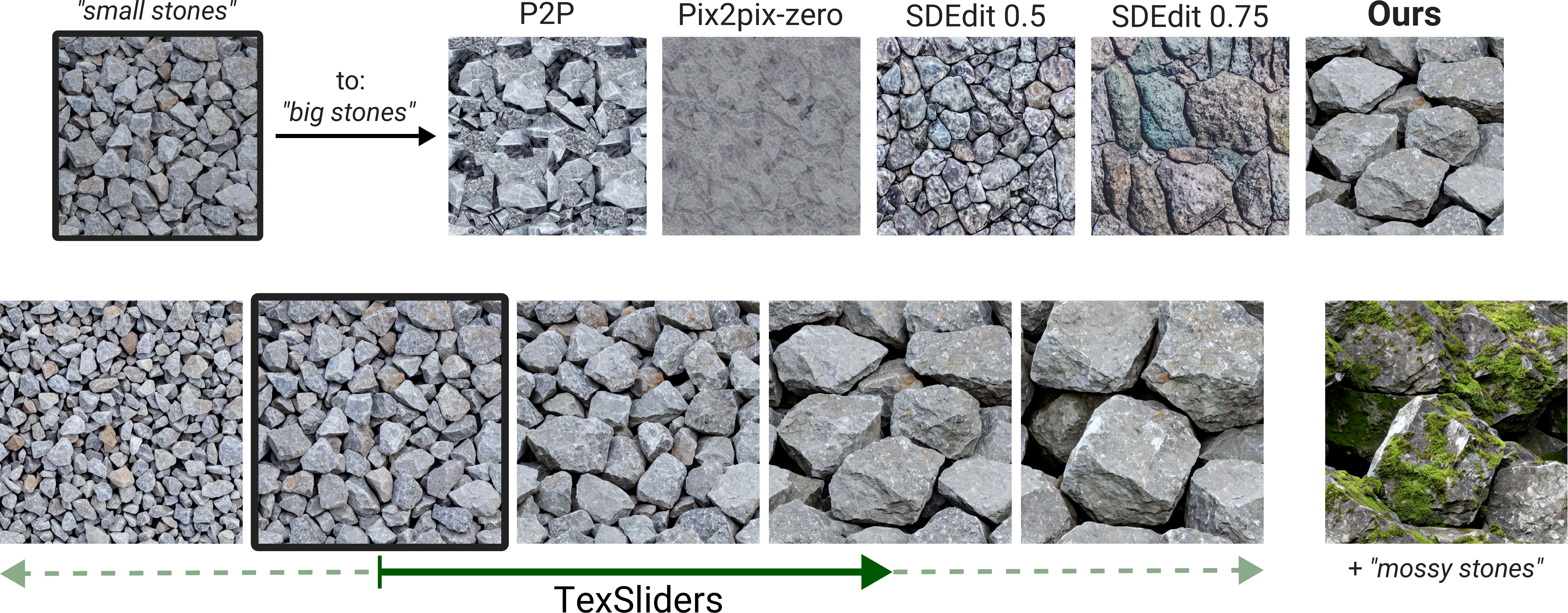}
\caption{We present TexSliders, a diffusion-based method for texture editing. \textit{Top row:} Given an input texture (highlighted in black) and a pair of natural language prompts describing an arbitrary edit (e.g., ``small stones'' to ``big stones''), existing methods fail to either perform the edit or preserve the texture identity. \textit{Bottom row:} Our approach finds a suitable editing direction in CLIP space~\cite{CLIP}, defining a \textit{slider} that allows to manipulate the texture along such direction (positive and negative) while preserving its original identity. Moreover, our method allows to combine multiple editing directions, as the rightmost image shows (``small stones'' to ``big, mossy stones'').
}
 \label{fig:teaser}
 \end{teaserfigure}

\maketitle

\renewcommand{\shortauthors}{J. Guerrero-Viu et al.}

\section{Introduction}

Generative models have given us unprecedented ability to create and edit visual content using natural language alone. Text-to-image diffusion models are trained on hundreds of millions of images collected online, and usually offer great results for the kind of images found at scale on the Internet~\cite{rombach2021stablediff, ramesh2022unCLIP}. However, using these models for domain-specific applications, such as texture generation, can pose new challenges. 

In this work, we target the creation of textures using diffusion models, focusing specifically on texture editing. Textures are ubiquitous in image manipulation, graphic design, illustrations, rendering, and 3D modeling. While diffusion models can occasionally produce a surprisingly good texture out of the box, more precise texture manipulation remains a challenge.
Recent work~\cite{hertz2022prompt,parmar2023zero} proposed to leverage \textit{attention maps} for general diffusion-based image editing, providing ``weights'' to focus the editing on the desired region. While this produces impressive photograph editing results, attention maps are not nearly as informative in the context of texture. We demonstrate in the evaluation that this limits their value for texture editing. 

We explore a different solution: texture manipulations in the CLIP~\cite{CLIP} embedding space, akin to latent manipulations in GANs~\cite{Patashnik_2021_ICCV}. We demonstrate our method on various \textit{editing directions} for textures such as weathering, scale, roughness, and more. Our approach allows to define new sliders for custom concepts with simple text prompts in a matter of minutes.

Our goal is to design a method with the following desirable properties: (1) the editing directions should be easy to define, (2) editing should preserve texture identity, (3) no ground-truth annotated data should be necessary, and (4) no modifications to the original diffusion model weights should be required. 

In our work, we consider texture ``identity'' to include all attributes of an input texture except the ones being edited.
To make the editing direction as easy as possible to define, we use two unrestricted textual prompts.
 We have experimented mostly with antithetic ones (e.g., ``aged wood'' to ``new wood'') and more specific ones (e.g., ``metal'' to ``rusty metal''), but any prompts that define an editing direction could be used.
We compute dozens of possible CLIP image embeddings for each of the two text prompts using a texture prior network~\cite{ramesh2022unCLIP, aggarwal2023Backdrop}. This gives us two clusters of embeddings, one for each prompt.
We then compute the direction between the two cluster centers. Averaging over multiple CLIP image embeddings factors out identity from the editing attribute to a large extent, but we still observe some identity variation when taking steps along the editing direction. We empirically find that many dimensions do not contribute to the attribute that is being edited, but rather contain noise that results in these identity variations. Thus, we propose to automatically select a subset of the dimensions from the average direction based on per-dimension intra- and inter-cluster statistical distances, effectively improving identity preservation even for large step sizes.

A key finding during our work is that identity preservation is significantly easier in diffusion models trained on image embeddings (as opposed to trained on text embeddings). Intuitively, this is due to the loss of information that occurs when describing an image (or a texture) with text, compared to the actual information contained in the image itself. We therefore leverage a diffusion model trained with image embeddings for our editing method. This diffusion model is moreover guaranteed to always produce tileable textures.

In summary, we propose an editing approach for textures by manipulating CLIP image embeddings for a pre-trained image-conditioned diffusion model using a texture diffusion prior. This is enabled by the following contributions:
\begin{itemize}
    \item Definition of CLIP-space editing directions from text prompts.
    \item Pruning of dimensions for improved identity preservation.
    \item Analysis of our editing directions' generalization, combinations, and use for generated images and real photographs.
\end{itemize}

Our method allows text-based texture editing
through sliders constructed from simple natural language prompts, while preserving texture identity, high image quality, and tileability.
\section{Related Work}
\paragraph{Diffusion-Based Generation}
Diffusion models~\cite{rombach2021stablediff,ramesh2022unCLIP} have recently revolutionized the field of image generation by creating high-quality image content from natural language prompts. They are based on the concept of diffusion probabilistic models~\cite{sohl2015diffusionbasis, song_score-based_2021, ho_ddpm_2020, nichol_ddpm2_2021} and typically employ a UNet architecture that performs iterative denoising from random per-pixel initial noise. For efficiency, this can be done in a latent space of reduced dimensionality~\cite{rombach2021stablediff}. Importantly, they also use a conditioning mechanism to guide the denoising from an input embedding~\cite{dhariwal_ddpm3_2021,nichol_glide_2021}. This embedding can be a class~\cite{dhariwal_ddpm3_2021}, a text prompt~\cite{nichol_glide_2021}, or an image~\cite{ramesh2022unCLIP}. Text and image prompts are typically encoded into vector embeddings by large vision-language models like CLIP~\cite{CLIP}. 
UnCLIP~\cite{ramesh2022unCLIP} (the base architecture of DALL-E 2) uses image embeddings previously generated by a prior network from text embeddings. \citet{aggarwal2023Backdrop} show how to specialize this prior for domain-specific contexts, such as textures. In our work we address the problem of text-based, continuous texture \textit{editing}, leveraging a latent diffusion model with a domain prior. Other works focus on texture generation on 3D shapes~\cite{richardson2023texture, chen2023text2tex}.

\paragraph{Diffusion-Based Image Editing}
Diffusion models are also widely applied to solve image editing problems. SDEdit \cite{meng2022sdedit} allows text-based image editing by leveraging a pretrained diffusion model, re-inserting noise and continuing diffusion towards a new condition. However, for larger edits, this approach cannot preserve the identity of the original image well, due to its stochastic nature.
Recently, methods like Prompt-to-Prompt~\cite{hertz2022prompt}, InstructPix2Pix~\cite{brooks2023instructpix2pix}, Pix2Pix-Zero~\cite{parmar2023zero}, or Diffusion Self-Guidance~\cite{epstein2023selfguidance} have shown how manipulating cross-attention maps can help retain the structure of the original image. These maps could also be used in combination with self-attention and certain intermediate layers of the diffusion UNet to better preserve global consistency~\cite{hertz2022prompt,tumanyan2023plug}.
These methods can be applied to diffusion-generated images or to real photographs with the help of inversion~\cite{mokady2023null}. However, cross-attention maps rely on semantic
information, which is usually missing in textures (see Figure~\ref{fig:crossattention}), and we find that the performance of these methods on texture editing degrades.
Imagic~\cite{kawar2023imagic} better preserves identity by fine-tuning both the text embedding and the weights of the diffusion model. While this gives compelling editing results, it requires a lengthy fine-tuning process for each target edit and each input image.
Other approaches have been proposed to guide the editing by example images instead of text~\cite{subrtova2023diffanalogies}.
Concurrently, Concept Sliders~\cite{gandikota2023sliders} proposes to find edit directions in the parameter space of a diffusion model through composable low-rank parameter updates.
While we share a similar goal, we use the orthogonal approach of finding edit directions in the CLIP embedding space, rather than in the parameter space of the diffusion model, and our method does not require any fine-tuning or other changes to the diffusion model itself.

\paragraph{Finding Directions in Generative Models}
Our work finds semantically meaningful directions in the latent space created by CLIP embeddings, which we leverage for texture editing. Exploring latent spaces of generative models aiming to disentangle semantic directions has been widely studied in the case of GANs~\cite{harkonen2020ganspace,wu2021stylespace}. This is more challenging for diffusion models, which typically lack such manageable space. Recent work has shown that a more convenient space can be found in pre-trained diffusion models through optimization~\cite{kwon2022diffusionlatent}. In contrast, our approach does not require optimization, simply requiring two text prompts to define a direction.

\paragraph{Texture Generation and Editing}
Texture synthesis is a long standing challenge in computer graphics.
In the last decade various deep learning approaches were proposed, targeting the synthesis of larger versions of input textures (e.g.,~\cite{Isola2017ImagetoImageTW, WCT-NIPS-2017, Henzler20, rodriguez2019automatic, Ulyanov2016TextureNF, Zhou2018NonstationaryTS, Shaham2019SinGANLA, mardani2020}). 
A popular approach is procedural modeling where the textures are defined as a combination of noise, patterns, and filters functions~\cite{SubstanceDes, pptbf, hu2022inverse}. One key benefit of such representation is controllability: as each function is defined by a set of parameters, this representation lets artists expose semantic control as sliders. Procedural content is however challenging to author, requiring significant artistic and technical skills, since the parameters do not always correspond to intuitive concepts, and their interactions may be exceedingly complex to understand \cite{Serrano2016}. 
Our approach targets the creation of similar sliders, enabling disentangled control over existing textures with simple text prompts, without the need for complex procedural representations.
\section{Diffusion-Based Texture Editing}
\label{sec:method}

\noindent In this section, we first show how practices that have led to great success in diffusion-based image editing are not well-suited to the particular case of texture manipulation. These observations motivate our choices when proposing a new method for texture editing, which we build upon an existing latent diffusion model without the need to re-train or fine-tune for editing directions.

\subsection{Preliminary Observations}
\label{subsec:obs}
We review design decisions made when editing with diffusion models, and analyze their effectiveness in the particular case of textures.

\begin{figure}[t]
 \centering
 \includegraphics[width=\columnwidth]{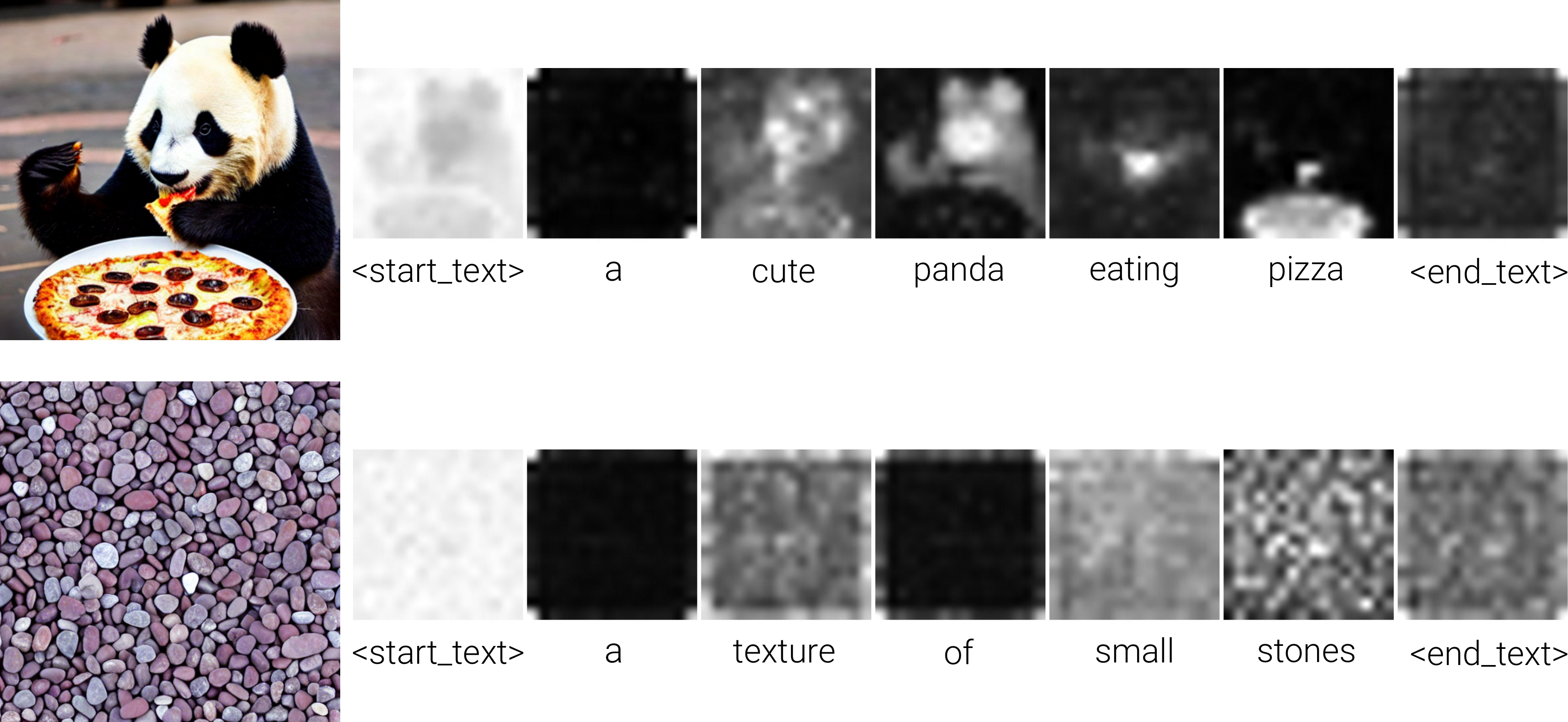}
 \caption{\textbf{Visualization of cross-attention maps}. We show maps at the last diffusion step of SD 1.4~\cite{rombach2021stablediff}, given two different input prompts. \textit{Top:} ``a cute panda eating pizza'' (non-texture). \textit{Bottom:} ``a texture of small stones'' (texture). The attention maps contain interesting semantic information for the panda image, but fail to capture the texture structure.}
 \label{fig:crossattention}
 \end{figure}
\begin{figure}[t]
 \centering
 \includegraphics[width=\columnwidth]{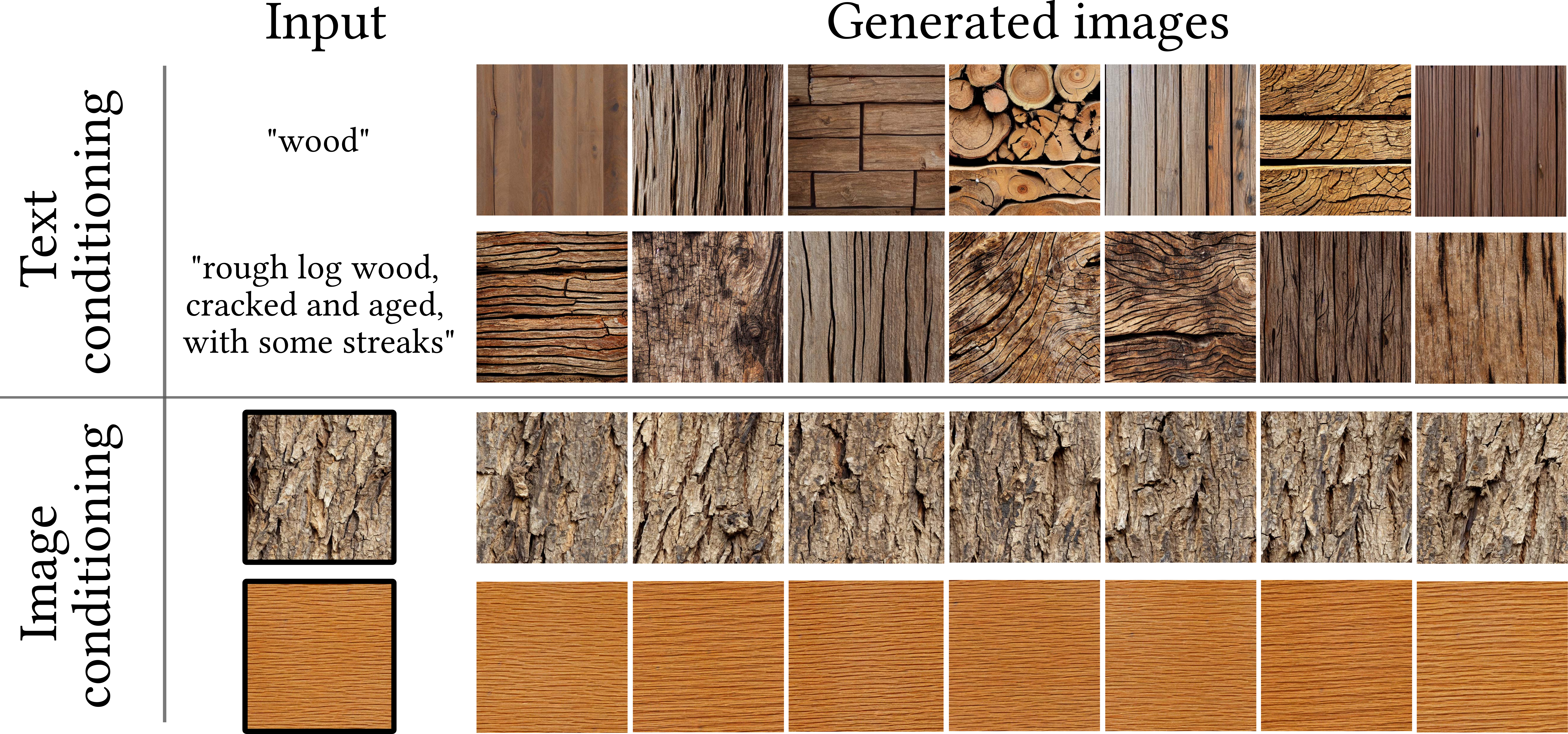}
 \caption{
 \textbf{Text vs. image conditioning in diffusion models.}
 Images generated by a diffusion model conditioned with two text prompts (top rows) and images (bottom rows).
 In both cases, each column represents a different seed when sampling the diffusion model.
Text conditioning, even with a specific prompt (second row), maps to a larger region in appearance space and can thus result in many different visual identities, while image conditioning maps the result to a more specific appearance.
Text conditioning is done with SD 1.4~\cite{rombach2021stablediff}, and image conditioning is done with the Latent Diffusion Model of~\citet{aggarwal2023Backdrop}.
 }
 \label{fig:text_vs_image_embeddings}
 \end{figure}
 \begin{figure*}[t]
 \centering
 \includegraphics[width=0.9\textwidth]
 {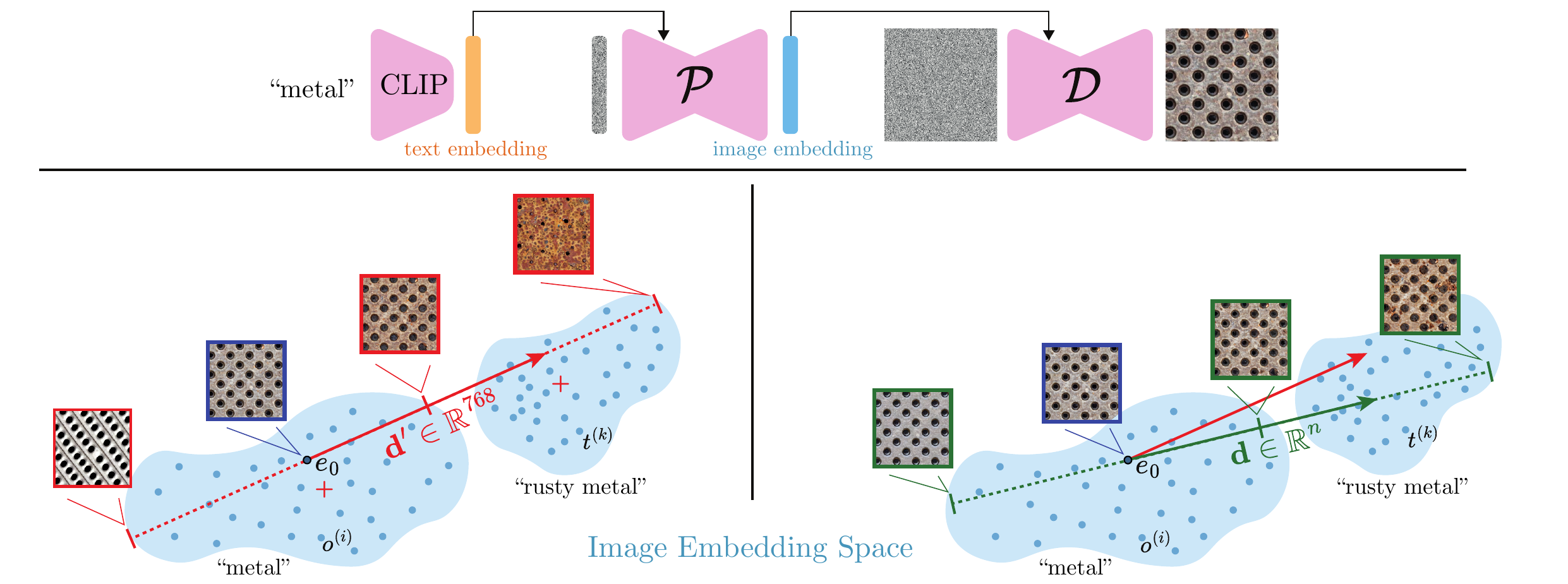}
 \caption{\textbf{Overview of our diffusion-based texture editing approach.} 
 \textit{Top row:} Our approach leverages a diffusion prior model $\mathcal{P}$ to convert text embeddings to image embeddings, enabling the use of an image-conditioned pre-trained diffusion model $\mathcal{D}$. \textit{Bottom row:} To perform the desired edits, we first compute direction $\mathbf{d'}\in \mathbb{R}^{768}$ (red arrow) as the difference between the centroids of the clusters (two small red crosses) formed by the image embeddings of the two prompts that define the edit (e.g., ``metal'' to ``rusty metal''). Naively applying this direction to a specific texture $e_0$ (highlighted in blue), leads to significant identity variations as we march along such direction (left). Instead, we select a subset of $n$ relevant dimensions ($n < 768$) that do contribute to the desired edit, leading to our final editing direction $\mathbf{d}$ (green arrow), which  preserves the identity of the input texture (right). We represent the high-dimensional CLIP image embedding space in 2D for visualization purposes. }
 \label{fig:method}
 \end{figure*}
\paragraph{Cross-Attention Maps}
Previous work on image editing with diffusion models (Prompt-to-Prompt~\cite{hertz2022prompt} and subsequent follow-up works~\cite{mokady2023null,brooks2023instructpix2pix}) made the key observation that the spatial layout of the generated image depends on the \emph{cross-attention maps}, which are spatial attention maps computed for each token of the text prompt (see Figure~\ref{fig:crossattention}, top row). These maps help with identity preservation during the editing process, avoiding the need to mask a part of the image as often done in previous methods~\cite{nichol_glide_2021,avrahami2022blendedEditing,avrahami2023blended}, and which is impractical for textures.
Cross-attention maps yield impressive results for non-texture images that typically have a clear structure with individual objects that correspond to phrases of the prompt. Attention maps encode this structure and can be used to both manipulate or preserve it.
However, textures usually lack such a clear separation into individual objects, and thus cross-attention maps are unable to capture structure and map it to the textual prompt,  failing to properly represent identity (Figure~\ref{fig:crossattention}, bottom row). Consequently, our approach does not employ cross-attention maps; instead, it relies on finding a direction in CLIP embedding space that preserves identity, as explained in Section~\ref{subsec:method}.

\paragraph{Domain-Specific Generation}
\sloppy
Large pre-trained models have yielded remarkable performance on specific subspaces when fine-tuned on smaller datasets~\cite{gal2022oneword,ruiz2023dreambooth,deschaintre2023visual}.
An alternative to fine-tuning an entire diffusion model is to attain domain-specific generation by means of a prior model trained on the specific data of interest~\cite{ramesh2022unCLIP,aggarwal2023Backdrop,zhou2023shifted}. This prior model, given a textual input, outputs an embedding (an image representation) that can then be used as conditioning to a diffusion decoder to produce the final image, constrained to the subdomain the prior was trained with. 

Textures are a specific subdomain within the larger distribution of images typically learnt by diffusion models. To help preserve identity and to constrain the generation to textures, our approach employs a diffusion prior model trained on textures, described in Section~\ref{subsec:model}, instead of fine-tuning the entire model. 

\paragraph{Manipulations in Image Embedding Space}
We seek to control the editing process by means of sliders with semantic meaning, which can be defined with simple text prompts. While the editing directions could thus be defined in text embedding space~\cite{ramesh2022unCLIP,deckers2023manipulating}, we observe that 
the notion of texture identity is more easily preserved in \textit{image} embedding space. Intuitively, it is easier to define the appearance of a texture with images than with textual descriptions, since these typically cannot describe all the details that constitute its identity. 
An intuition of this is given in Figure~\ref{fig:text_vs_image_embeddings}, where we compare text- and image-conditioned generation, and show that even a detailed textual prompt leads to a much higher variance in the generated images than using image conditioning.

\subsection{Diffusion Model}
\label{subsec:model}
We use a Latent Diffusion Model alongside a \emph{domain diffusion prior}~\cite{ramesh2022unCLIP,aggarwal2023Backdrop} trained for the texture domain, as described in Section~\ref{subsec:obs}.

The domain diffusion prior is a diffusion model trained to generate CLIP \emph{image} embeddings matching a given CLIP \emph{text} embedding, as shown in Figure~\ref{fig:method} (top). Note that this is a generative process, as there are generally multiple image embeddings matching a text embedding. More specifically, the diffusion prior is trained to generate image embeddings in the ``texture'' part of the CLIP L/14 embedding space, encouraging the generation of texture-like images from the main, pre-trained latent diffusion model. The diffusion model is trained on a dataset of 77M images with no humans or text, and the prior is trained separately on a 10M subset of that dataset, classified as textures.
The prior further allows the use of text prompts to interact with a network trained on image embeddings, while retaining high generation quality and prompt alignment.

\subsection{Computing Sliders for Texture Editing}
\label{subsec:method}
As described above, our method employs a diffusion model $\mathcal{D}$ coupled with a texture diffusion prior $\mathcal{P}$, and performs the editing in image embedding space. 
We thus seek to define a \textit{direction} $\mathbf{d}$ in such space, specified by a pair of simple text prompts that describe the original and target appearance (e.g., from ``metal'' to ``rusty metal''), and which will act as a slider: marching along such direction (positive and negative) will progressively increase or decrease the intensity of the desired edit. 

Figure~\ref{fig:method} (bottom) illustrates this. We first compute the CLIP text embeddings 
of the original and target prompts, 
and feed them to the prior $\mathcal{P}$, yielding image embeddings 
within the texture domain that fit the textual descriptions. In order to obtain a robust representation of the editing prompts, we compute a set of $n_e$
image embeddings for the \textit{original} and \textit{target} prompts by sampling the diffusion prior. We term these image embeddings $\mathbf{o}^{(i)}$ and $\mathbf{t}^{(k)}$, respectively, with both $i$ and $k$ $\in \{1..n_e\}$.
The number $n_e$ of image embeddings is an adjustable parameter that we set to 150 in our experiments for both the original and target embeddings. We then define an initial editing direction $\mathbf{d'}$ in image embedding space as the difference between the centroids of the clusters formed by the original and target embeddings. Note that $\mathbf{d'} \in \mathbb{R}^{768}$, as it corresponds to the dimensionality of the image embedding obtained from the diffusion prior model. 

Each component $d'_j$ is given by:
\begin{equation}
    d'_j = \frac{1}{n_e} (\sum_k t_j^{(k)} - \sum_i o_j^{(i)}) \, .
\end{equation}

Computing multiple image embeddings to obtain this initial direction aids in disentangling the relevant attribute(s) from the rest but does not suffice, leading to poor results (see Figure~\ref{fig:method}, bottom left).
To better preserve the identity of the input texture while progressively changing only the desired attribute, we need to select the subset of relevant dimensions, avoiding those that do not contribute to the desired edit, or lead to identity variations.

We select the relevant dimensions based on their intra-cluster variability, as given by the standard deviation $std$, compared to their inter-cluster variability, as given by the distance between cluster centroids.
Dimensions with high inter-cluster variability typically contribute more to the desired edit, while dimensions with high intra-cluster variability typically encode the identity of each individual texture. We therefore select those dimensions whose inter-cluster variability is larger than the intra-cluster one, as those are more likely to be representative of the edited attribute only.
We set the remaining dimensions to zero. The components of our resulting direction vector $\mathbf{d}$ (shown in Figure~\ref{fig:method}, bottom right) are thus:
\begin{equation}
    d_j = \left\{
    \begin{matrix}
    d'_j \, , & \textup{if} \; 
    |\tilde d'_j | > \tau \cdot \underset{k}{std}(\tilde t_j^{(k)}) \; \textup{and} \; 
    |\tilde d'_j | > \tau \cdot \underset{i}{std}(\tilde o_j^{(i)}) \\ 
    0 \, , & \textup{otherwise.} \hfill
   \end{matrix}
   \right.
\end{equation}
The relationship is modulated by a threshold $\tau$ (typically 0.8 in our experiments), and applied over normalized vectors $\mathbf{\tilde d'}$, $\mathbf{\tilde t}$ and $\mathbf{\tilde o}$, so that the comparison is meaningful. 

Given $\mathbf{d}$, we can march along that direction to obtain different degrees of the desired edit, for instance by means of a slider. In particular, given the image embedding $\mathbf{e}_{0}$ 
of a texture we wish to edit, the final image embedding $\mathbf{e}_{\alpha}$ 
becomes:
\begin{equation}
\mathbf{e}_{\alpha} = \mathbf{e}_{0} + \alpha \cdot \mathbf{d} ,
\label{eq:slider}
\end{equation}
where $\alpha$ modulates the intensity of the edit, and can take positive or negative values. The resulting
$\mathbf{e}_{\alpha}$
is then used as conditioning in the diffusion model $\mathcal{D}$ to generate the final, edited texture image.

\paragraph{Definition of New Sliders.} Besides using pre-computed sliders for editing, 
users can easily create new ones adapted to their needs by providing two text prompts; defining a new slider takes about 2 minutes on a single A10G GPU (in our unoptimized implementation).
\section{Results}
\label{sec:results}

We evaluate our diffusion-based editing approach on textures representing different kinds of  materials, with various editing directions. We first explain the details of our test dataset containing both diffusion-based generated and real photographs of textures (Section~\ref{subsec:dataset}). We qualitatively evaluate our results on \textit{generated} images in Section~\ref{subsec:qualitative_results}, including an ablation study of the main components of our method, and additional results on real \textit{photographs} of textures in Section~\ref{subsec:real_results}. In Section~\ref{subsec:sota}, we compare to alternative diffusion-based image editing approaches, including a quantitative comparison to previous works. Finally, Section~\ref{subsec:generality} includes further evaluation about the generality of our method for multiple practical applications.
Additional results and implementation details are available in the Supplemental Material.

\begin{figure*}
\includegraphics[width=0.89\textwidth]{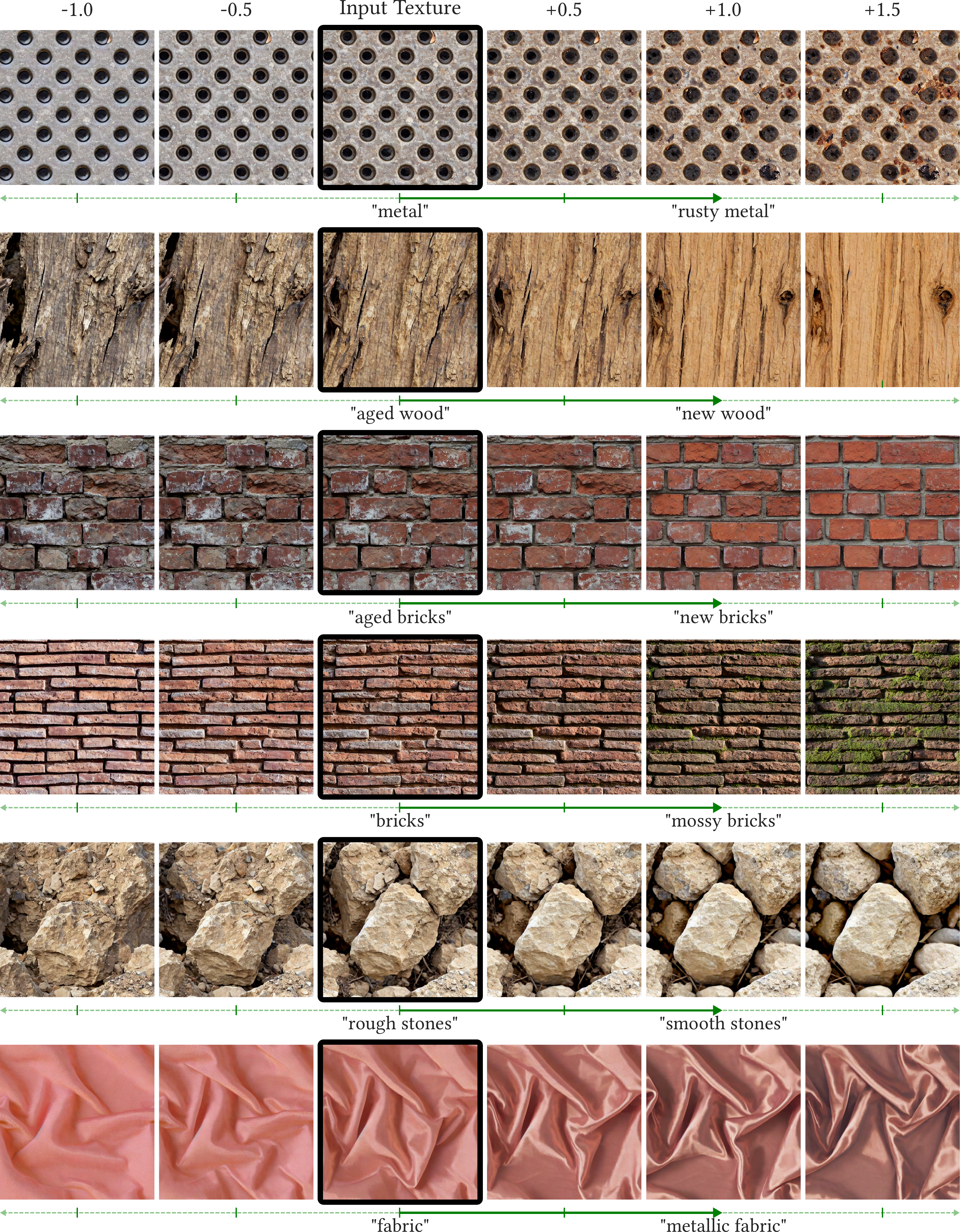}
\caption{\textbf{Qualitative results.} We show our method on different kinds of materials for various editing directions. We can see that our method applies convincing editing, including when extrapolating the directions (two leftmost columns and rightmost column), and preserves well the texture identity. Here we use generated input textures; Figure~\ref{fig:real_images} and the Supplemental Material provide results also on photographed inputs.}
\label{fig:qualitative_results}\end{figure*}

\subsection{Test Dataset}
\label{subsec:dataset}
To exhaustively evaluate our approach on a diverse set of textures, and given the lack of standard benchmarks for this purpose, we create a test dataset of generated and real photographs of textures, paired with editing attributes. 
We first define a set of seven material types, typically found in 3D creation tools:
\emph{bricks}, \emph{fabric}, \emph{leather}, \emph{metal}, \emph{paper}, \emph{stones}, and \emph{wood}. Then, we define between one and four prompts for nine of the main editing attributes that people use to describe materials~\cite{deschaintre2023visual}, such as ``rough'' and ``smooth'' for the attribute \emph{touch}, or ``aged'' and ``rusty'' for \emph{weathering}.
We filter the \{material, attribute\} pairs that are semantically valid, and generate two different textures for each pair, using the original prompt formed by material and attribute (e.g., ``smooth stones'', or ``aged wood'') and two different random seeds for the diffusion prior. Finally, we add two real photographs of textures for each material type (from Adobe Stock~\shortcite{AdobeStock}), making a total of 80 images (66 generated and 14 real) associated to a pair of original and target prompts. Our full dataset is included in the Supplemental Material.

\subsection{Qualitative Evaluation}
\label{subsec:qualitative_results}
Our editing method is conceptually simple and zero-shot, since it does not require ground-truth data, nor training or fine-tuning of the diffusion model.
We evaluate its performance on our test dataset, and show some representative results in Figure~\ref{fig:qualitative_results}. Since the output of our pipeline is a \emph{direction} in the embedding space that conditions the diffusion model, we can obtain different degrees of the desired edit by marching along the direction (varying $\alpha$ in Eq.~\ref{eq:slider}), even extrapolating in the forward ($\alpha > 1.0$, rightmost column) and backward ($\alpha < 0.0$, first and second columns) directions.
Our method successfully performs challenging edits, such as those related to weathering (``aged'', ``rusty'', or ``mossy''), metallicness, or touch (``rough'' to ``smooth''), on various material types. 
Importantly, most of these edits require globally changing all regions of the texture, which we accomplish while preserving the identity of the input texture.
\begin{figure}
\begin{tabular} {cccc}
  & \hspace{-3.5mm} $-$  \hspace{-1mm} & \hspace{-5.5mm} $\leftarrow$ Input Texture  $\rightarrow$  & \hspace{-5.5mm} \hspace{-1mm}   $+$ \vspace{2mm} \\
\vcentered{ \begin{tabular} {@{}c@{}} \small{Single image} \\ \small{embedding} \\ $n_e = 1$ \end{tabular}} & \hspace{-3.5mm} \includegraphics[width=0.12\textwidth]{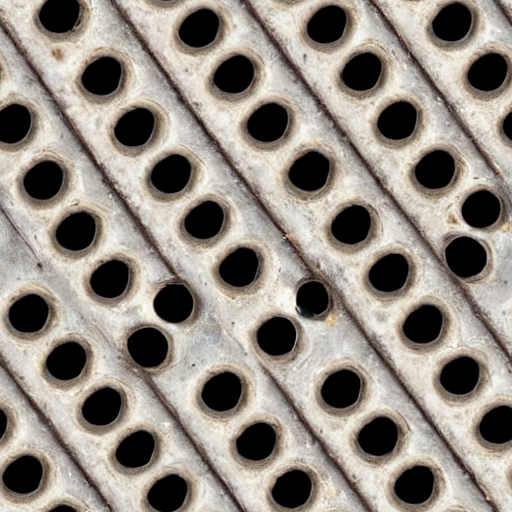} & \hspace{-5.5mm} \includegraphics[width=0.12\textwidth, frame=1pt]{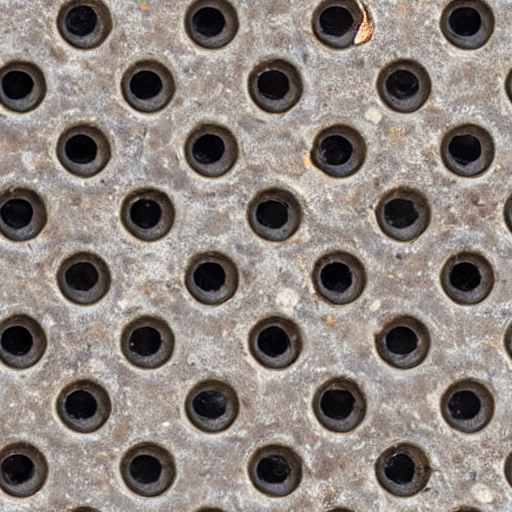} & \hspace{-5.5mm} \includegraphics[width=0.12\textwidth]{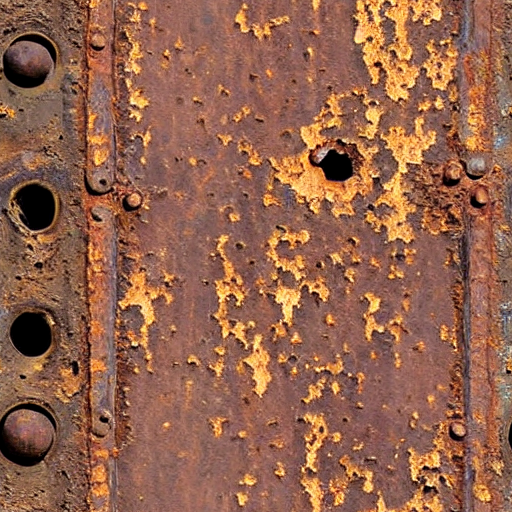} \vspace{1mm} \\
\vcentered{ \begin{tabular} {@{}c@{}}  \small{All} \\ \small{dimensions} \\ $\mathbf{d'}$ \end{tabular}} & \hspace{-3.5mm} \includegraphics[width=0.12\textwidth]{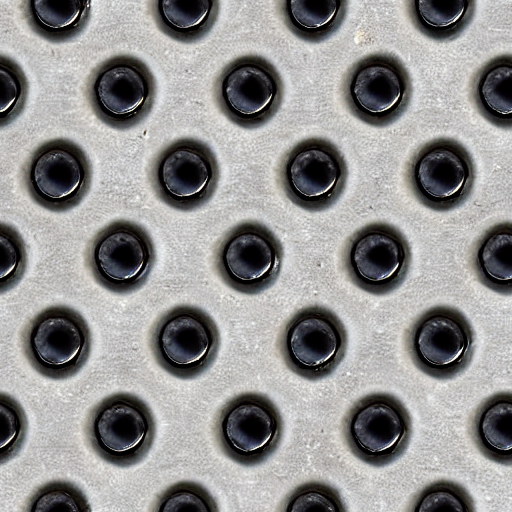} & \hspace{-5.5mm} \includegraphics[width=0.12\textwidth, frame=1pt]{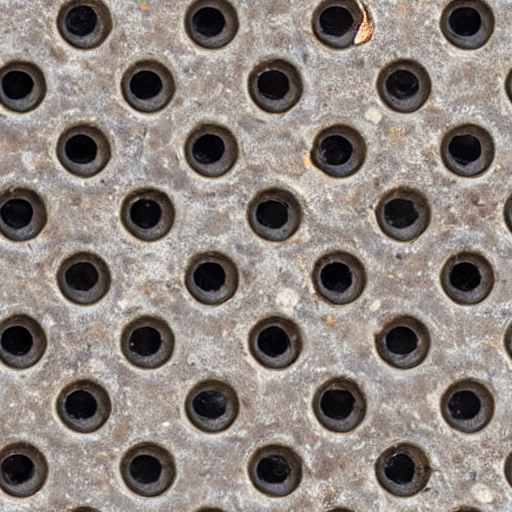} & \hspace{-5.5mm} \includegraphics[width=0.12\textwidth]{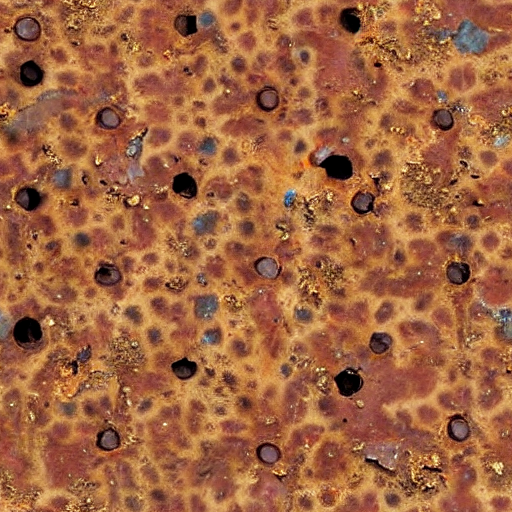} \vspace{1mm} \\
\vcentered{{ \begin{tabular} {@{}c@{}} \small{Ours}  \\ ($\tau = 0.8$) \\ $\mathbf{d}$ \end{tabular}}} & \hspace{-3.5mm} \includegraphics[width=0.12\textwidth]{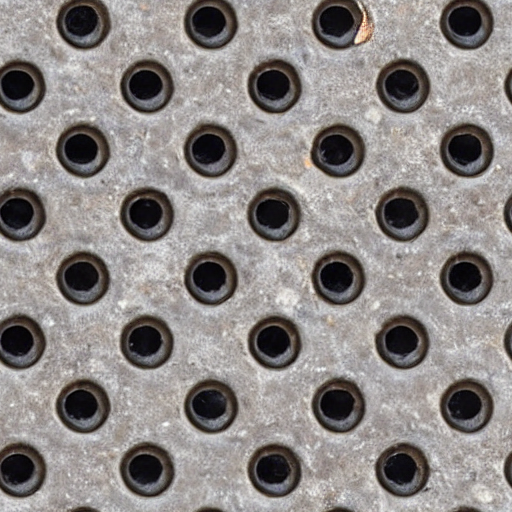} & \hspace{-5.5mm} \includegraphics[width=0.12\textwidth, frame=1pt]{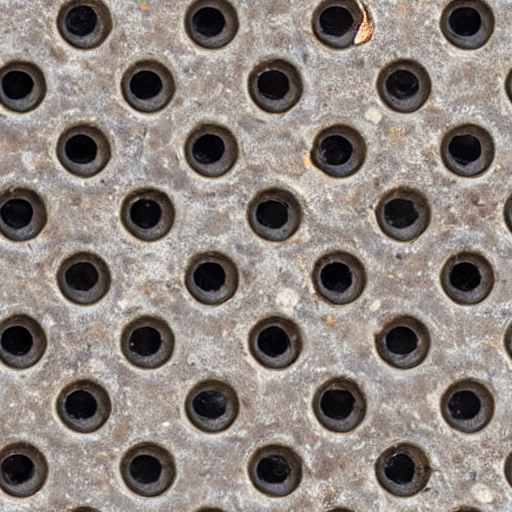} & \hspace{-5.5mm} \includegraphics[width=0.12\textwidth]{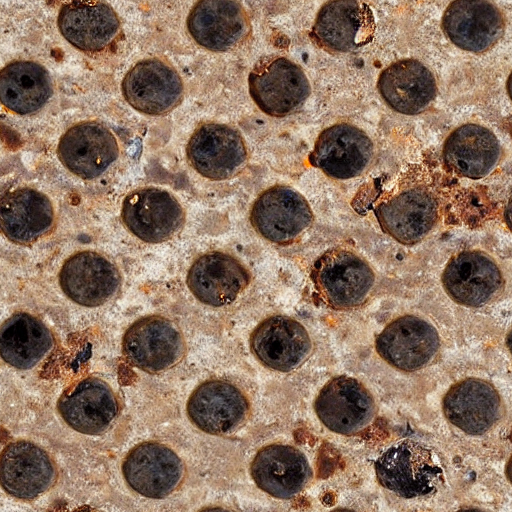} 
\end{tabular}
\caption{\textbf{Ablation study}. We show qualitative ablations for the direction ``metal'' to ``rusty metal'' (positive and negative). We compare to using a single image embedding for original and target prompts ($n_e = 1$, top row) and to using direction $\mathbf{d'}$, computed for multiple image embeddings ($n_e = 150$), but including all the 768 dimensions of the direction (middle row). Both options show a higher identity shift than our approach (bottom row), which successfully edits the texture maintaining its original structure.
}
\label{fig:ablations}
\vspace{-0.75cm}
\end{figure}
\begin{figure*}
\includegraphics[width=\textwidth]{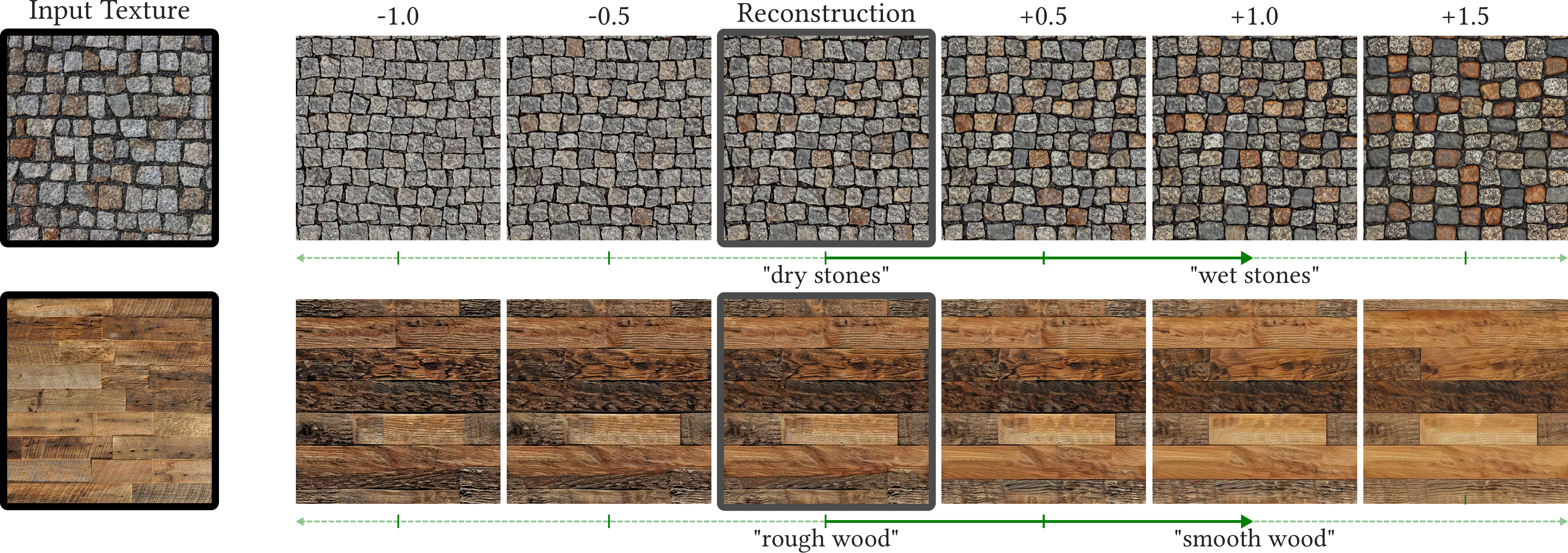}
\caption{\textbf{Photographed texture edits.} We use the CLIP embedding of the input texture to reconstruct the photograph using the diffusion model, and apply our editing approach to the reconstructed texture (highlighted in grey), showing that our approach can be used for non-synthetic textures as well.}
\label{fig:real_images}
\end{figure*}

\paragraph{Ablation Study.}
\label{subsec:ablation}
We ablate our key design choices and evaluate how they impact performance, in terms of disentanglement of the editing attributes and identity preservation.
Figure ~\ref{fig:ablations} shows results for the ``metal'' to ``rusty metal'' example. In the top row, we use a \textit{single} image embedding given the original and target prompts ($n_e = 1$); the results fail to preserve the identity of the input texture, due to the entanglement of attributes along the resulting direction.
The middle row shows results using several image embeddings, but using the direction $\mathbf{d}'$, i.e., including \textit{all} dimensions in CLIP space ($\mathbf{d}' \in \mathbb{R}^{768}$). Although the editing improves, the identity of the input texture is still not properly maintained. In contrast, our approach (bottom row) selecting a \textit{subset} of relevant dimensions yields good edits while preserving the identity. Further ablations, including different methods to select the subset of relevant dimensions, can be found in the Supplemental Material.

\subsection{Real Photographs}
\label{subsec:real_results}
Our results are not limited to diffusion-based generated textures. Since the diffusion model is conditioned on CLIP image embeddings, and these are a faithful representation of the texture appearance (see Figure~\ref{fig:text_vs_image_embeddings}), we can use the CLIP embedding of any input image as conditioning to reconstruct it. From this embedding and a pair of prompts, we can use our method to compute the editing direction $\mathbf{d}$ and generate textures with different degrees of editing.
We show examples of these results using real photographs in Figure~\ref{fig:real_images}, where our method successfully edits different material types and attributes, such as wetness and smoothness. 
While out of the scope of this work, we could further improve the accuracy of the reconstruction, and hence our editing results, by inverting the diffusion model (e.g., using null-text inversion~\cite{mokady2023null}).

\subsection{Comparison to the State of the Art}
\label{subsec:sota}
Last, we compare our approach to previous state-of-the-art diffusion-based image editing methods. To the best of our knowledge, ours is the first diffusion-based editing framework targeting the domain of textures, so we compare to general methods for natural images. 
Since these methods use Stable Diffusion~\cite{rombach2021stablediff} as generative model, and do not include a textures domain prior, for fair comparison, we add the preamble ``a texture of...'' to all prompts in our test dataset. With that, we restrict the generated images of Stable Diffusion to the textures subspace. We use the same number of diffusion steps, $50$, and the same classifier-free guidance (guidance scale = $7.5$) for all methods. 
We include results of SDEdit~\cite{meng2022sdedit}\footnote{ Public implementation in the \emph{diffusers} library: \url{https://huggingface.co/docs/diffusers/api/pipelines/stable_diffusion/img2img}} at two different noise levels ($0.5$ and $0.75$), Prompt-to-Prompt (P2P)~\cite{hertz2022prompt}\footnote{ Official implementation paired with null-text inversion from: \url{https://github.com/google/prompt-to-prompt/}}, and Pix2pix-zero~\cite{parmar2023zero}\footnote{ Official implementation from: \url{https://github.com/pix2pixzero/pix2pix-zero}}, the last two using cross-attention maps to keep structural information. \\

\begin{figure*}
\includegraphics[width=0.9\textwidth]{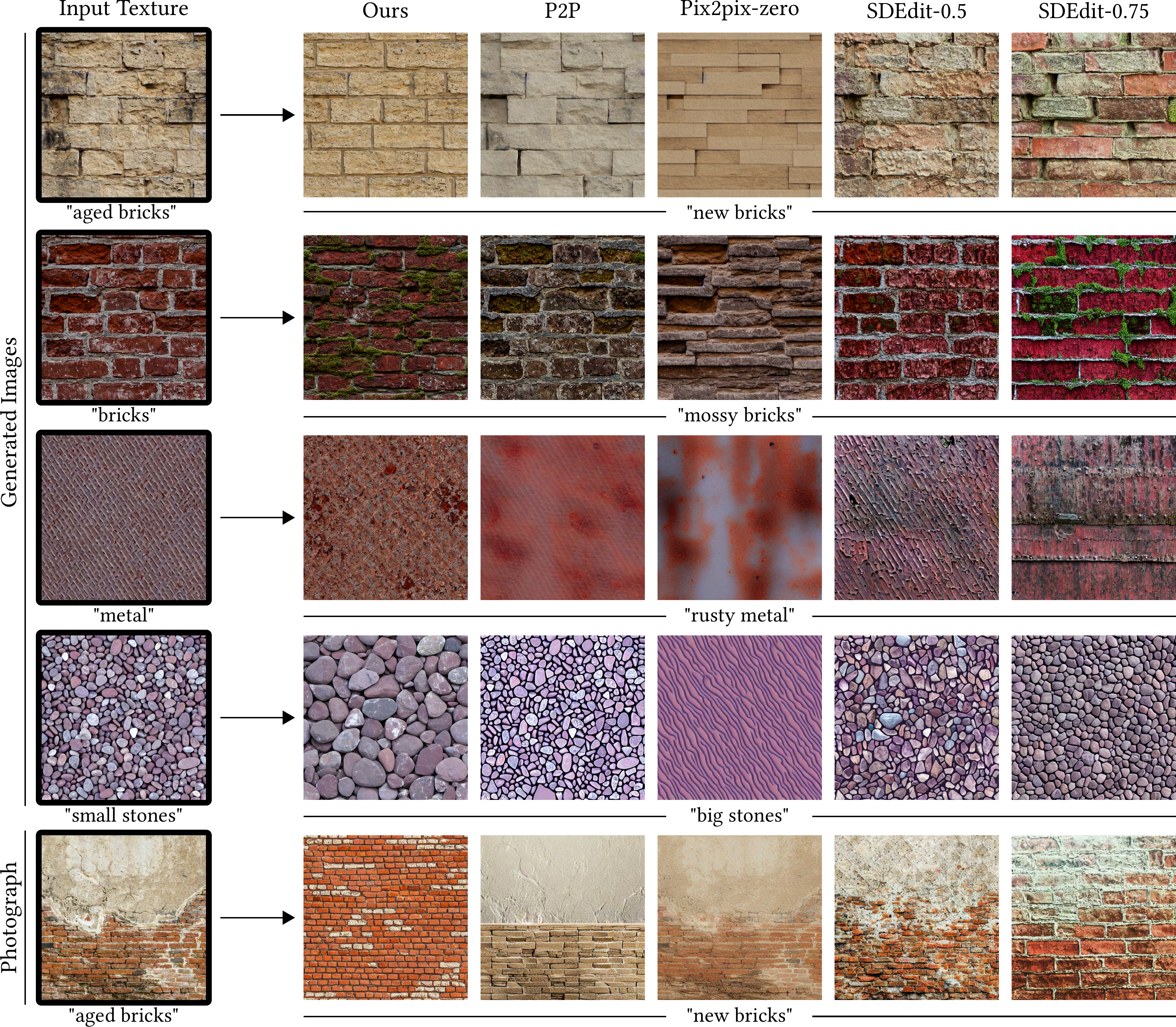}
\caption{\textbf{Comparisons to state-of-the-art methods.} For every input image, and pair of prompts that define an editing direction, we show the output of the following diffusion-based editing methods (from left to right): Ours ($\tau=0.8$, $\alpha=1.5$), Prompt-to-Prompt (P2P)~\cite{hertz2022prompt} (cross\_replace\_steps $=0.8$, self\_replace\_steps $=0.3$), Pix2pix-zero~\cite{parmar2023zero} (xa\_guidance $=0.5$), and SDEdit~\cite{meng2022sdedit} with noise levels $0.5$ and $0.75$, respectively. We can see that our approach leads to much better identity preservation and realistic editing, following the desired direction.}
\label{fig:comparisons_sota_1.5}
\end{figure*}

As shown in Figure~\ref{fig:comparisons_sota_1.5},
our results exhibit both strong identity preservation,
as well as good adherence to the target prompt. 
Contrarily, SDEdit shows an undesirable trade-off between both, where if using a low noise level ($0.5$) the identity is preserved but the appearance is hardly edited (e.g., ``mossy bricks''), or if the noise level is increased to $0.75$ the edit is performed but the identity changes noticeably (e.g., ``rusty metal''). Pix2pix-zero and P2P both demonstrate clear loss of identity in several cases, or even some noticeable visual artifacts that push the result out of the texture domain (e.g., ``big stones''), due to their reliance on
cross-attention maps that are unsuitable for texture editing.
We also outperform previous methods on real photographs (last row), despite the fact that our approach does not use inversion for reconstruction, unlike P2P and Pix2pix-zero.
Remarkably, our approach is the only one able to successfully perform scale changes in the pattern (e.g., ``small stones'' to ``big stones''), as in that case texture identity is not defined by the spatial structure.

\begin{table}[]
    \caption{\textbf{Quantitative metrics} (mean values on our full test dataset, including generated images and photographs). Our approach's results better match the required edit direction (CLIP-Direction) than previous work, while also better preserving the identity of the input texture (CLIP-Im2Im).}
    \centering
    \begin{tabular}{|c|c|c|}
    \hline
     Method & $\uparrow$ CLIP-Direction & $\uparrow$ CLIP-Im2Im \\
      \hline
     P2P &  0.0880 & 0.9033\\
     Pix2pix-zero & 0.0811  & 0.8667\\
     SDEdit 0.5 & 0.0706 & 0.8753\\
     SDEdit 0.75 &  0.0690 & 0.8707\\
     Ours & \textbf{0.1063} & \textbf{0.9303}\\
     \hline
    \end{tabular}
    \label{tab:quantitative_results}
    \vspace{-0.55cm}
\end{table}
%

\paragraph{Quantitative Analysis.}
\label{subsec:metrics}
As a perceptual task, it is not trivial how to quantitatively measure the quality of an edit (both in terms of adherence to the edit and identity preservation). In order to give an objective comparison, we follow previous works and use metrics in CLIP embedding space, as follows:
\begin{itemize}
\item CLIP-Direction: Measures edit adherence as cosine similarity between the directions in CLIP space of the pre/post edit text prompts and images~\cite{gal2022stylegan, brooks2023instructpix2pix}.
\item CLIP-Im2Im: Measures identity preservation as cosine similarity in CLIP space between input and edited image~\cite{brooks2023instructpix2pix}.
\end{itemize}

We show in Table~\ref{tab:quantitative_results} that our approach significantly outperforms previous work in both metrics, supporting our conclusions from the qualitative comparison.
\begin{figure*}
\includegraphics[width=\textwidth]
{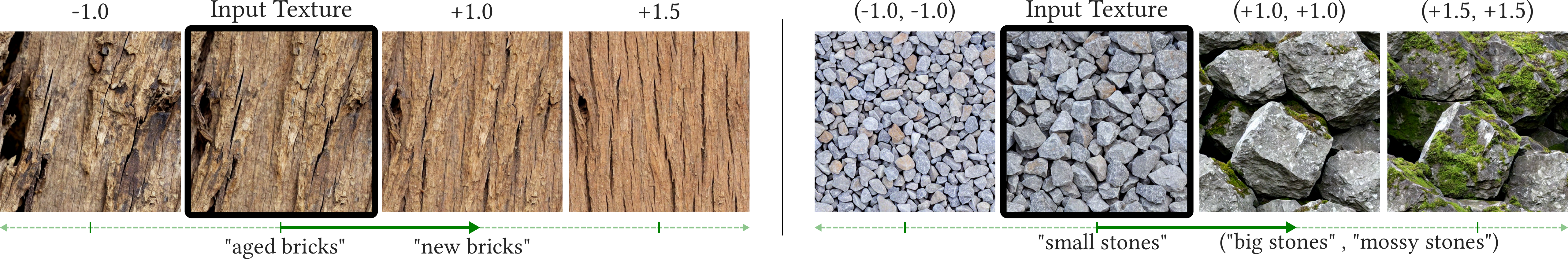}
\caption{\textbf{Generalization and multi-dimensional edits.} \textit{Left:} We show that using directions computed for bricks (``aged'' $\rightarrow$ ``new'') generalize to other classes such as wood. \textit{Right:} We evaluate the combination of multiple editing directions (``small stones'' $\rightarrow$ ``big stones'' and ``mossy stones''). We show here the diagonal of the matrix of edits, where both directions are changed by the same amount (see complete matrices of edits and additional results in the Supplemental Material).}
\label{fig:additional_experiments}\end{figure*}

\subsection{Generality}
\label{subsec:generality}
We further evaluate our approach's generality along three axes:
(a) generalization of edit directions across material types, (b) combination of multiple edit directions, and (c) using images instead of text to describe the editing direction. 

To evaluate (a), we apply directions computed for a certain material to a different one, showing reasonable results in exchange for a small identity loss (see Figure~\ref{fig:additional_experiments}, left, and the Supplemental Material). We demonstrate (b) through full matrices of edit combinations in the Supplemental Material.
The diagonal of the matrix that combines scale and mossiness for stones (from (-1,-1) to (+1.5,+1.5)) is shown in Figure~\ref{fig:additional_experiments}, right.
Last, regarding (c), when an editing direction cannot be captured well by two textual prompts but we have instead a set of image pairs that illustrate the change, we can first obtain the CLIP embeddings of such set of images and then apply our method to compute the editing direction. This is possible because our approach operates in CLIP image space, and generalizes well across multiple material types. We show successful results using $20$ images of textures cropped and resized to define both original (``zoomed in'') and target (``zoomed out'') clusters in the Supplemental Material.
\section{Discussion and Limitations}
We propose TexSliders, a diffusion-based texture editing approach that enables slider-like disentangled editing for customized directions. To do this, we leverage image-conditioned diffusion models together with a texture diffusion prior and explore the CLIP image space to derive identity-preserving editing directions. We show that our simple yet effective solution is superior to other image editing methods when applied to the texture domain. 

\begin{figure}[t]
 \centering
 \includegraphics[width=\columnwidth]{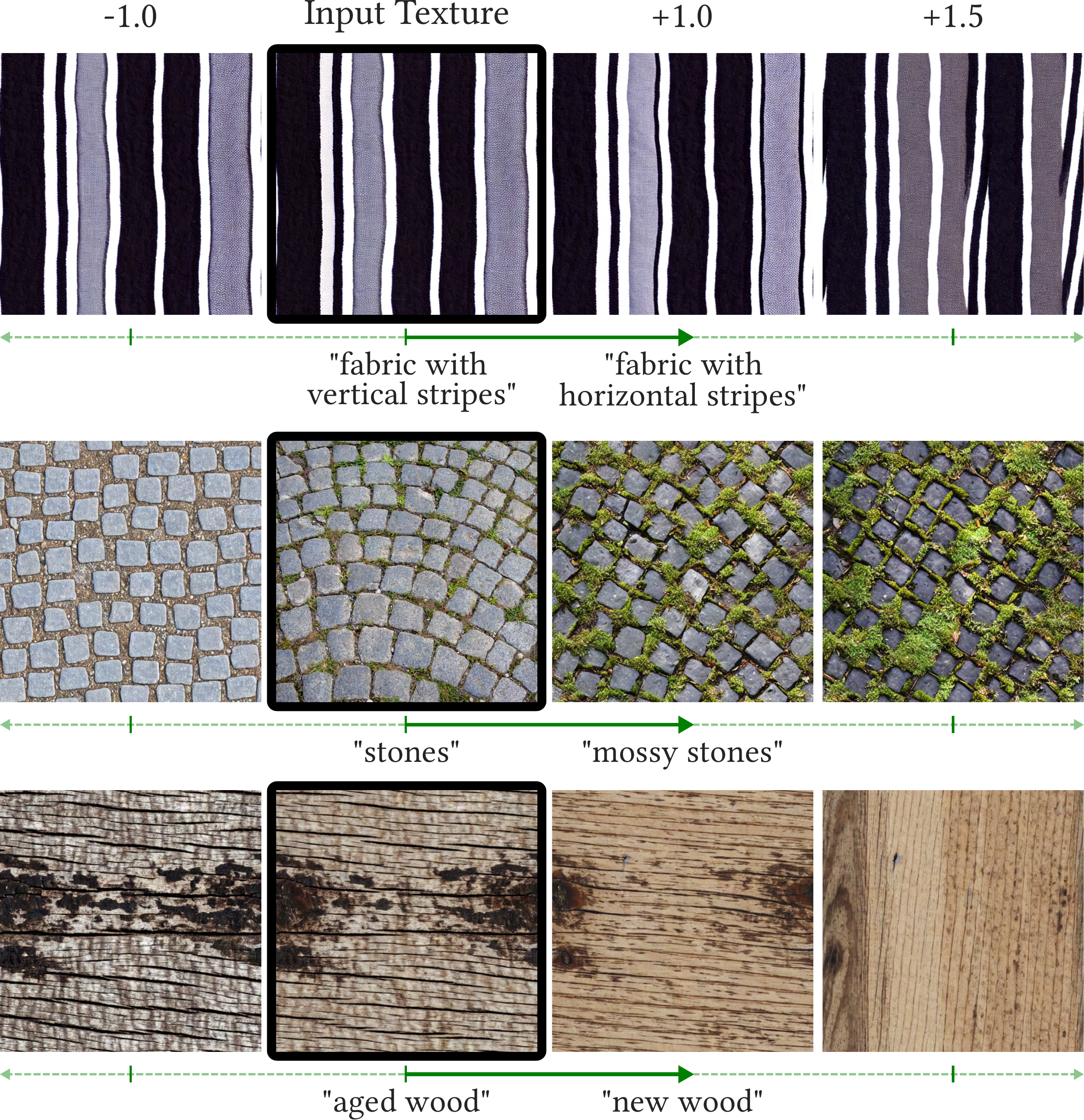}
 \caption{\textbf{Limitations.} \textit{Top}: Biases in CLIP and the diffusion model, such as an existing bias towards vertically-oriented patterns, can lead to failed editing. \textit{Middle}: The identity of an input texture (e.g., the arrangement of the cobblestones) may be altered due to remaining noise in the editing direction. \textit{Bottom}: Textures may not be guaranteed to preserve identity when extrapolated too far from the input image, leading to embeddings that are outside the sampled CLIP distributions. }
 \label{fig:limitations}
 \end{figure}

Our method is not free of limitations. Since we leverage CLIP space for diffusion-based texture editing, our approach inherits some limitations from the diffusion model and the CLIP representation. For some cases, the intended editing may fail to change the texture appearance significantly, and a key observation in this regard is that CLIP and the diffusion model can be more sensitive to some concepts compared to others. For example, we found that CLIP represents the concept ``vertical'' better than ``horizontal''; as a result, the CLIP-conditioned generative model biases the generated textures toward vertically-oriented patterns, and textures with vertical stripes can struggle to turn horizontal despite following the corresponding editing direction (Figure~\ref{fig:limitations}, top row). We also note that in some cases the identity of the input texture is not perfectly preserved during the editing; an example is shown in Figure~\ref{fig:limitations} (middle row), where the edit is performed, but the pattern of the cobblestones is not completely preserved. This may be due to noise or spurious dimensions remaining present in the editing direction, and other methods for dimension selection could be explored. Identity preservation can also be hampered when extrapolating in the editing direction too far from the input texture, leading to embeddings outside the sampled distribution (e.g., last column in Figure~\ref{fig:limitations}, bottom row).

One must note that the notion of identity in textures is not a formally defined one, even if it is arguably easy to judge qualitatively. For our purposes, identity depends on the input texture and the attributes being edited, and it includes all other attributes of the input texture except the ones being edited. We leave an in-depth study of texture identity in diffusion-based models for future work. We focus on color textures in this work, from which material maps could be approximated~\cite{controlmat,MatFusion}. Extending our editing approach to such material maps should be possible, although current material diffusion models rely on a relatively limited amount of data (orders of magnitude smaller than for images), which would limit their expressivity. Better understanding the CLIP embedding space for images~\cite{zhou2023clippae}, and for the particular case of textures~\cite{wu2022CLIPtexture, deschaintre2023visual}, is also an interesting line of research,
both for synthesis and editing. Finally, the integration of our approach within a complete user interface would allow for artists to provide feedback about its usability and desirable improvements.
\\

Our method enables pre-computation of editing directions that can be used as sliders to edit specific attributes of any input texture. Besides, one can compute any new custom direction specified via two unrestricted text prompts---or alternatively with sample images, see Section~\ref{subsec:generality}---in under two minutes.
Our work is not only the first to propose diffusion-based editing for the domain of textures, but also has the potential to be adapted to other domains, such as vector graphics-like images.
\vspace{1cm}
\begin{acks}
This work has been partially supported by grant PID2022-141539NB-I00, funded by MICIU/AEI/10.13039/501100011033 and by ERDF, EU, and by the European Union’s Horizon 2020 research and innovation programme under the Marie Skłodowska-Curie grant agreement No. 956585 (PRIME).
Julia Guerrero-Viu developed part of this work during an Adobe internship, and was also partially supported by the FPU20/02340 predoctoral grant.
We thank Daniel Martin and Sergio Izquierdo for insightful discussions and help preparing the final figures, and Ajinkya Kale for insightful discussions, as well as the team that developed the internal backbone diffusion model. 
\end{acks}

\clearpage
\bibliographystyle{ACM-Reference-Format}
\bibliography{bibliography}

\end{document}